\documentclass[fleqn,usenatbib]{mnras}

\usepackage{fix-cm}
\usepackage{newtxtext,newtxmath}

\usepackage[T1]{fontenc}

\DeclareRobustCommand{\VAN}[3]{#2}
\let\VANthebibliography\thebibliography
\def\thebibliography{\DeclareRobustCommand{\VAN}[3]{##3}\VANthebibliography}

\usepackage{graphicx}	
\usepackage{amsmath}	

\usepackage{textcomp}
\usepackage{ulem}
\newcommand{\mic}{$\mu$m }  
\newcommand{\lred}{$\lambda_R (3.3)_{1/2}$ }
\newcommand{\lblue}{$\lambda_B (11.2)_{1/2}$ }
\newcommand{\jwst}{\textit{JWST} }
\newcommand{\iso}{\textit{ISO SWS} }
\newcommand{\nc}{$N_C$ }

\title[Examination of the PAH Hypothesis]{A Critical Examination of the PAH Hypothesis}

\author[A.T. Tokunaga et al.]{
Alan T. Tokunaga,$^{1}$\thanks{E-mail: tokunagaa001@gmail.com}
Lawrence S. Bernstein,$^{2}$
and Takashi Onaka$^{3}$
\\
$^{1}$Institute for Astronomy, University of Hawaii, 2680 Woodlawn Dr., Honolulu, HI  96822, USA\\
$^{2}$63 Forest Glen Ln, Topsham, ME 04086, USA\\
$^{3}$Department of Astronomy, Graduate School of Science, The University of Tokyo\\
7-3-1 Hongo, Bunkyo-ku, Tokyo 113-0033, Japan
}

\date{Accepted 2025 Oct. 28. Received 2025 Oct. 27; in original form 2025 Sept. 17}

\pubyear{\the\year{}}

\begin{document}
\label{firstpage}
\pagerange{\pageref{firstpage}--\pageref{lastpage}}
\maketitle

\begin{abstract}
The polycyclic aromatic hydrocarbon (PAH) hypothesis proposes that the aromatic infrared bands (AIBs) observed at 3.3, 6.2, 7.7, 8.6, 11.3, and 12.7 \mic originate from gas-phase PAH molecules. 
These bands exhibit consistent peak wavelengths and profiles in diverse sources, and ISO SWS and JWST spectra show a nearly identical red wing of the 3.3 \mic AIB and blue wing of the 11.2 \mic AIB in the dominant Class A sources.  
This spectral uniformity suggests that the AIBs arise from a small, well-defined set of gas phase PAH species, regardless of the excitation conditions or the nature of the source such as HII regions, reflection nebulae, planetary nebula, young stellar objects, or the diffuse interstellar medium. 
However, a small number of gas phase PAH species is inconsistent with current modeling of the AIBs that require a wide range of PAH types and sizes. 
It is also inconsistent with the lack of observed UV and optical absorption bands from gas phase PAH molecules. 
Furthermore, there is no plausible formation pathway to produce only a small number of specific PAH molecules in the interstellar medium.  
These issues require quantitative investigation in order to definitively establish gas-phase PAH molecules as the carrier of the AIBs.

\end{abstract}




\section{Introduction}



The PAH hypothesis postulates that the main Aromatic Infrared Bands (AIBs) observed in emission at 3.3, 6.2, 7.7, 8.6, 11.2, and 12.7 \mic arise from gas phase PAH molecules \citep{Leger84, Allamandola89, Peeters11, Peeters21}.  
Although it has explained many aspects of the AIB emission, no specific PAH molecules have been detected at UV, optical, or infrared wavelengths. 
The presence of PAH molecules is inferred by fitting the AIB spectrum with the weighted combination of individual PAH spectra calculated using the NASA Ames PAH database \citep[Ames PAHdb;][]{Boersma14,Bauschlicher18,Mattioda20}.  
These fits use assumptions regarding the temperature-dependent band locations and profile widths. 
Qualitative fits to the AIB spectrum can be achieved with combinations of neutral, cation, and anion PAH spectra, with the relative contributions varying depending on the source, excitation conditions, and degree of ionization  \citep[i.e.][]{Boersma18,Shannon19,Maragkoudakis20}.
A large number of PAHs are assumed since this averages out the individual spectral features and explains the absence of narrow UV and optical absorption bands from specific gas phase PAH species 
\citep{Clayton03, Gredel11, Steglich11, Steglich12}.
Modeling of the AIB emission using the Ames PAHdb typically uses 30-50 PAH species 
\citep{Andrews15, Bauschlicher18, Ricca24}.

The main AIB spectra show a lack of spectral diversity which is not expected under the PAH hypothesis. The main AIB peak wavelengths and widths are very similar in many diverse sources which includes HII regions, planetary nebulae, reflection nebulae, and the diffuse interstellar medium (ISM). 
\citet{Boulanger98} and \citet{Boulanger99} remarked on the similarity of the AIB spectra over a range of 10$^3$ in excitation intensity.  
The AIBs arise from the stochastic heating of molecules or very small grains (VSGs) and the composition of the emitting carrier is apparently very similar in diverse objects and locations within the Milky Way and other galaxies. 
The lack of spectral diversity of the dominant Class A\footnote{
The classification scheme of Class A, B, and C for the AIB wavelengths and profiles was introduced by \citet{Peeters02} and \citet{vanDiedenhoven04}. Class D was defined by \citet{Matsuura14}. 
} 
AIBs at 3-13 \mic was further discussed by \citet{Pech02}, \citet{Verstraete01}, \citet{Andrews15}, \citet{Tokunaga21}, \citet{Mackie22}, and \citet{Chown24}.
However a lack of spectral diversity is difficult to reconcile with gas phase PAH molecules, whose band positions and widths vary with excitation energy, molecular size, ionization, and molecular symmetry  (i.e. symmetric or irregular) \citep{Joblin95-1, Bauschlicher08, Bauschlicher09, Ricca12, Ricca24, Tokunaga21}. 

Addressing these issues has led to the idea that there may be only a small number of specific gas phase PAHs. 
\cite{Verstraete01} modelled the AIB spectrum of NGC 2023 and M17 SW and concluded that explaining the AIB emission with gas phase PAH molecules requires a limited range of gas phase PAHs and a mechanism to maintain a similar temperature distribution of the emitting PAHs in various excitation conditions. \cite{Pech02} modelled the spectrum of IRAS 21282$+$5050 and suggested that the gas phase PAHs have similar structures (i.e. a family of compact PAHs) or that the emitting molecules reach a “solid state limit” (a size-independent constant spectrum) in order to explain the similarity of the AIB spectra.
\citet{Andrews15} proposed a limited set of PAHs to explain the lack of spectral diversity of the AIBs while \citet{Mackie22} stated more strongly that `the interstellar PAH family is dominated by a few very stable PAHs', a conclusion that was also reached by \citet{Chown24}.   

Therefore conflict exists between the underlying assumptions of the PAH hypothesis that the AIB carrier consists of a large population of gas phase PAH molecules, while the observational evidence suggests that the AIB carrier involves only a small number of PAH molecules. 
In this paper, we critically examine the evidence underlying this conflict and the consequences for the PAH hypothesis.  
We focus on the 3.3 \mic CH stretch and 11.2 \mic out-of plane (OOP) bending modes since the wings in the spectrum of these bands strongly constrain PAH species which could account for the AIBs.

We use the term "AIB" to refer specifically to the main emission bands at 3.3, 6.2, 7.7, 8.6, and 11.2 \micron.  
We focus on the main AIBs and the clues they offer regarding the identification of the carrier. 
The discussion of the numerous minor AIB features identified by Chown et al. (2024) and Peeters et al. (2024) is outside the scope of this paper. 
We use the term "lack of spectral diversity" to refer to the similar emission profile of the AIBs, the nearly constant red wing of the 3.3 \mic AIB, and the nearly constant blue wing of the 11.2 \mic AIB. 
Section~\ref{sec:lack of diversity} presents observational evidence for the lack of spectral diversity. 
Section~\ref{sec:unresolved questions} examines the conflicts with the PAH hypothesis. 
Section~\ref{sec:other bands and the continuum} looks into unresolved questions concerning other emission features linked to the PAH hypothesis.
Section~\ref{sec:discussion} discusses our findings, and 
Section~\ref{sec:summary} provides a summary of our conclusions. 

\section{The AIB profiles}
\label{sec:lack of diversity}


In this section we discuss the red wing of the 3.3 \mic and the blue wing of the 11.2 \mic AIB bands, both of which remain remarkably constant in diverse astrophysical sources.  
We also discuss the 6.2, 7.7, and 8.6 \mic AIBs which show nearly constant peak wavelengths in diverse sources. 
Although the lack of spectral diversity was noted previously, \jwst observations combined with \iso spectra provide more quantitative insights about the nature of the PAHs responsible for the AIBs.

\subsection{The Red Wing of the 3.3 \texorpdfstring{$\mu$m}{micron} AIB}   
\label{sec:3.3 profile}

\begin{figure}
	\includegraphics[width=\columnwidth]{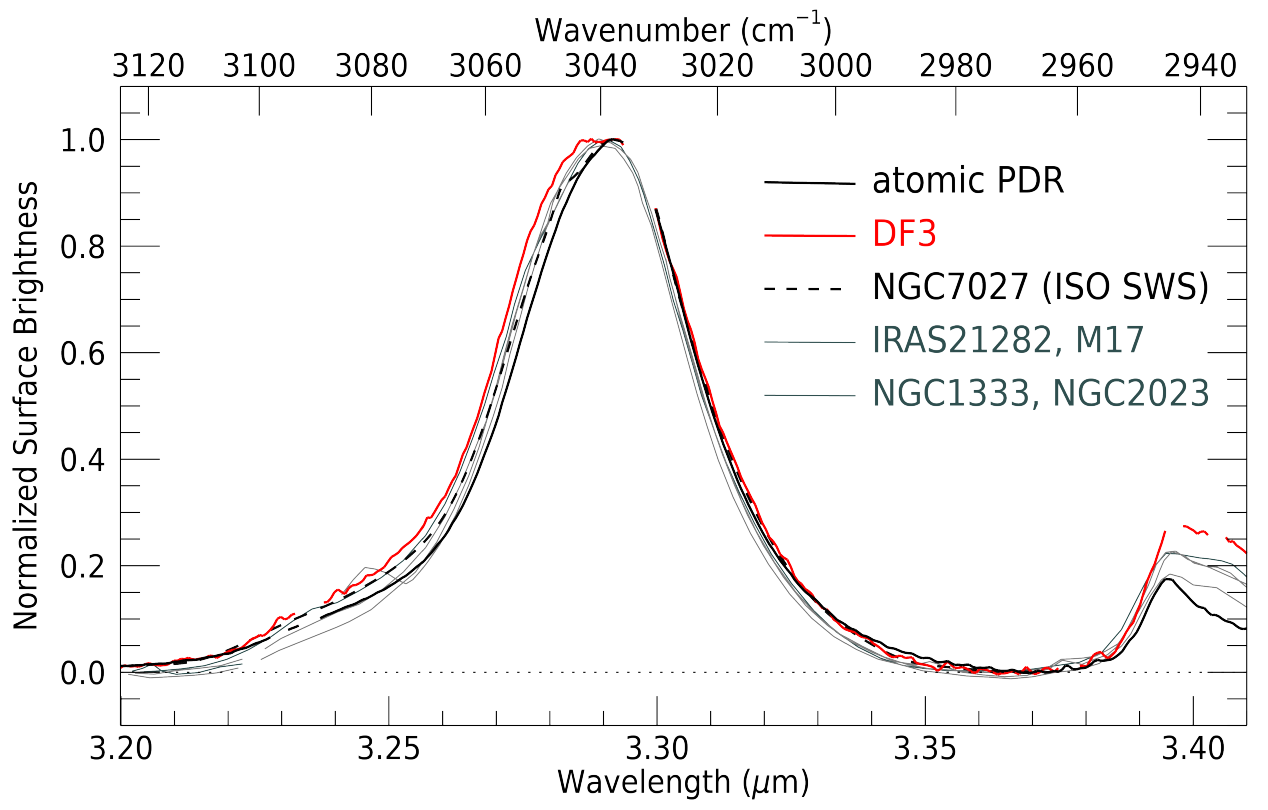}
    \caption{The normalized \textit{JWST} spectra of the atomic PDR region and DF3 regions of the Orion Bar (black and red solid lines, respectively) after subtraction of the continuum and plateau emission. 
    The units of the surface brightness are MJy sr$^{-1}$.
    For comparison the normalized \textit{ISO} spectrum of NGC 7027 from \citet{Tokunaga21} is shown (black dashed line). 
    The \textit{JWST} spectra of IRAS 21282, M17, NGC 1333, and NGC 2023 (grey solid lines) are from \citet{Boersma23}.  
    Note that the \textit{ISO} spectrum of NGC 7027 was 
    taken with a 14\arcsec $\times$ 20\arcsec\ aperture while the JWST spectra were taken with a 3\arcsec\ diameter aperture.  }
    \label{fig:3.3 profile}
\end{figure}

\begin{figure}
	\includegraphics[width=\columnwidth]{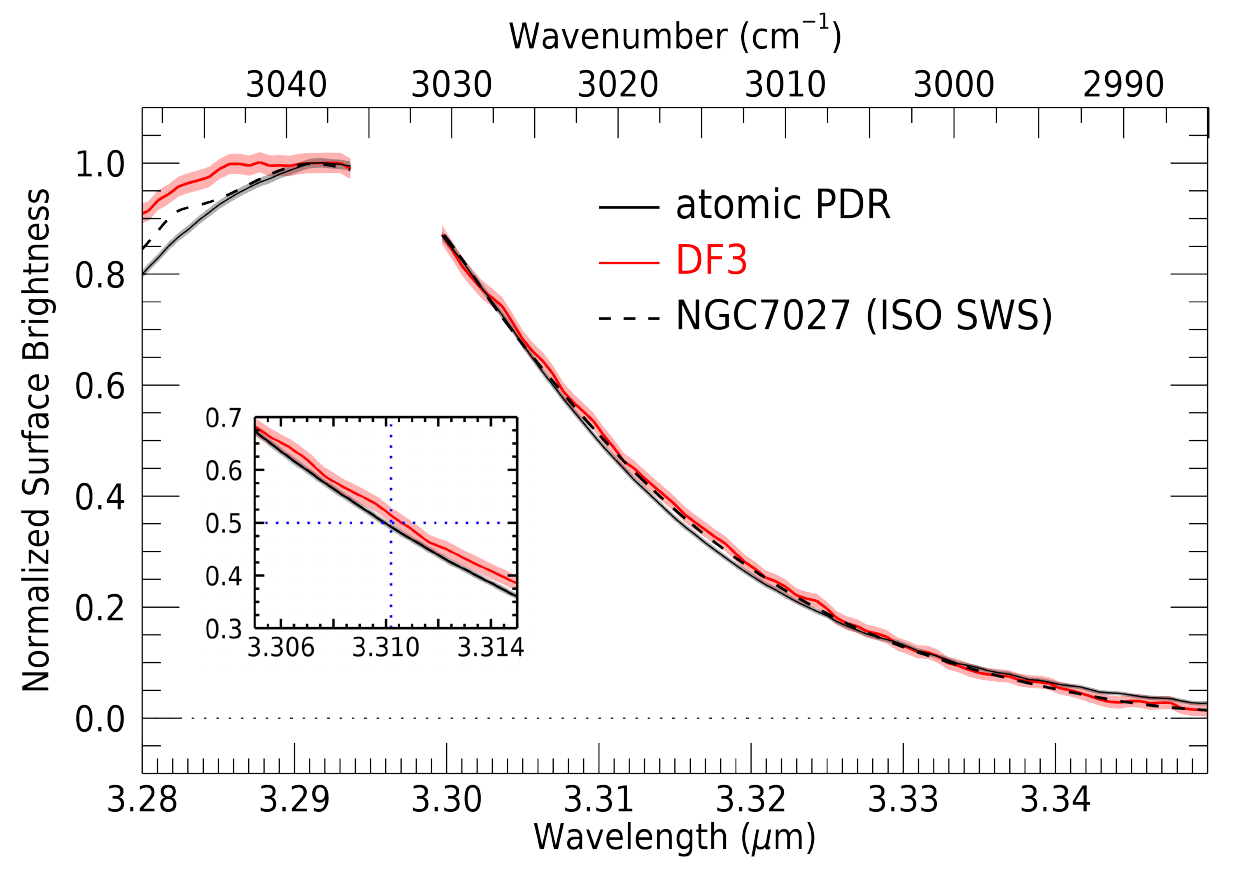}  
    \caption{ Expanded view of the red wing of the 3.3 \mic band.  The one sigma error bars of the atomic PDR and DF3 spectra are shown as shaded areas around each profile.  The inset shows that the error bars are overlapping and therefore $\lambda_R (3.3)_{1/2}$ for the atomic PDR and DF3 spectra is not significantly different within the uncertainty of the observations. 
    }
    \label{fig:expanded 3.3 profile}
\end{figure}

High signal-to-noise \textit{JWST} spectra of the atomic photodissociation region (PDR) and dissociation front 3 (DF3) regions in the Orion Bar were obtained from the PDRs4All archive\footnote{https://pdrs4all.org/seps/\#highly-processed-data-products}. 
The atomic PDR and the DF3 regions are defined by \citet{Chown24} and \citet{Habart24}.
The spectra normalized to the peak surface brightness are shown in Figures \ref{fig:3.3 profile} and \ref{fig:expanded 3.3 profile} along with the \textit{ISO SWS} spectra of other sources. 
In these spectra, a linear continuum and the 3.4 \mic plateau as described in \citet{vanDiedenhoven04} have been removed. 
In addition, the narrow emission lines have been removed.

The 3.3 \mic band profile and peak wavelength in the Orion Bar region are very similar in the atomic PDR and the DF3 regions even though the DF3 region has 
a UV field intensity that is 3 times lower than the atomic PDR region \citep{Goicoechea25}. 
The average wavelength at half peak intensity, $\lambda_R (3.3)_{1/2}$, of the red wing of the 3.3 \mic AIB is 3.3102 \mic with a small spread between the two spectra of 0.0007 \mic (0.64 cm$^{-1}$) as shown in the inset in Figure \ref{fig:expanded 3.3 profile}. 

\citet{Boersma23} presented \textit{JWST} spectra of M17, NGC 1333-SVS3, NGC 2023, and IRAS 21282+5050.
These objects are examples of an HII region, a young protostellar object, a reflection nebula, and a planetary nebula. 
We digitized the spectra\footnote{
WebPlotDigitizer, https://apps.automeris.io/wpd4/} 
shown in their Figure 10 to allow a comparison to the \textit{JWST} Orion Bar spectra.
This is shown in Figure \ref{fig:3.3 profile}. 
However, it is important to note that Boersma et al. employed a different method for subtracting the 3.4 \mic plateau emission than that of \citet{vanDiedenhoven04} and this paper.  
This could introduce small systematic differences in the band profile.

For the combined JWST and ISO SWS spectra, we find $\lambda_R (3.3)_{1/2}$ = $3.3094 \pm 0.0012$ \mic ($3021.72 \pm 1.14$ cm$^{-1}$). 
The uncertainty, indicated by `$\pm$', 
is the spread in $\lambda_R (3.3)_{1/2}$ and likely would be smaller if a consistent method had been used to subtract the linear and plateau components from all spectra shown in Figure \ref{fig:3.3 profile}. Despite the wide range of excitation environment and source types, the red wing is remarkably similar.  This uniformity suggests that a stable and well-defined set of PAH species may be responsible for the 3.3 \mic AIB, regardless of the local excitation conditions. This is discussed further in Section \ref{sec:3.3 AIB profile}.

\subsection{The blue wing of the 3.3 \texorpdfstring{$\mu$m}{micron} AIB}  
\label{sec:3.3 blue wing}

The blue wing shows a wider spread in the half-power point, with $\lambda_B (3.3)_{1/2}$ = 3.2690 $\pm$ 0.0019  \mic (3059.05  $\pm$ 1.8  cm$^{-1}$).  
As shown in Figure 1, the atomic PDR spectrum has the narrowest full width at half maximum (FWHM) and the DF3 region the broadest.  
\citet{Chown24}, \citet{Peeters24}, and \citet{Pasquini24}  suggest the UV-shielded DF3 region contains a higher abundance of smaller and more fragile PAHs compared to the atomic PDR region.  
For a given UV photon that is absorbed, the smaller PAHs in the DF3 region would reach a higher temperature than larger PAHs in the atomic PDR region, which could explain the broader FWHM observed in the DF3 region.  
This explanation does not, however, give a reason for the nearly constant \lred and the very similar overall profile of the 3.3 \mic AIB. 
This consistency implies that a very similar set of PAHs is present in both the atomic PDR and DF3 region, as well as in the other sources shown in Figure 1.  

\subsection{The Blue Wing of the 11.2 \texorpdfstring{$\mu$m}{micron} AIB }   
\label{sec:11.2 blue wing}

\begin{figure}
	\includegraphics[width=\columnwidth]{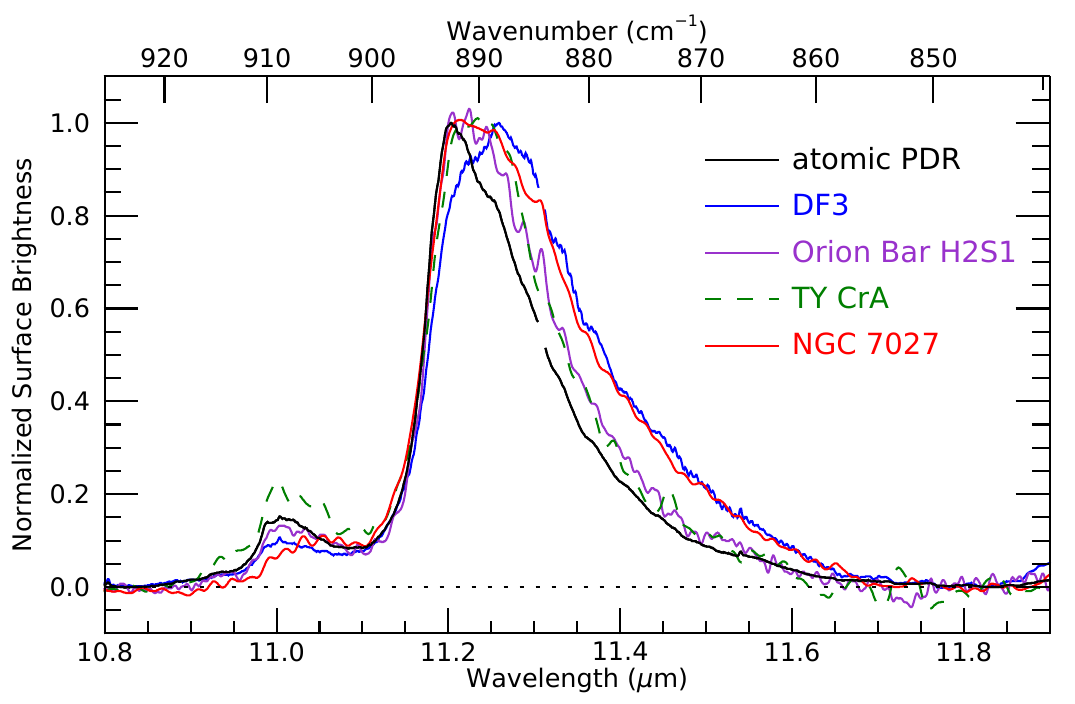}  
    \caption{ Comparison of the normalized 11.2 \mic AIB profile in the \textit{JWST} atomic PDR and  
    DF3 regions of the Orion Bar to the \textit{ISO SWS} spectra of the Orion Bar H2S1, TY CrA, and NGC 7027. 
    Emission lines have been removed.  
    In the PAH hypothesis, the weaker emission band at 11.0 \mic is attributed to PAH cations.
    }
    \label{fig:11.2 profile}
\end{figure}

Figure \ref{fig:11.2 profile} shows a comparison of the \textit{JWST} and \textit{ISO SWS} spectra of the 11.2 \mic band. 
The \textit{JWST} atomic PDR and DF3 spectra of the Orion Bar region were obtained from the processed spectra in the PDRs4All archive. 
We used the multiplicative version of the MIRI spectra.  
The ISO SWS spectra of the Orion Bar H2S1 (TDT 69501806), Herbig Ae/Be star TY CrA (TDT 33400603), and the planetary nebula NGC 7027 (TDT 33800505) spectra were obtained from the spectral archive of \citet{Sloan03}, where the SWS Target Dedicated Time (TDT) identification number is given in parentheses. 

The continuum was estimated using a spline fit following \citet{Hony01}, and this was subtracted from the spectra to produce the 11.2 \mic band profiles shown in Figure \ref{fig:11.2 profile}. 
The atomic PDR spectrum is Class A and the DF3 spectrum is Class B.
Compared to the atomic PDR region, the \iso spectra have a broader FWHM compared to the atomic PDR region.   The broader spectra may have resulted from the large \iso aperture size (14\arcsec $\times$20\arcsec) 
which likely encompasses a mixture of Class A and Class B emission regions.  The higher angular resolution spectra of \jwst might reveal distinct Class A and B regions within the \iso aperture.

\begin{figure}
	\includegraphics[width=\columnwidth]{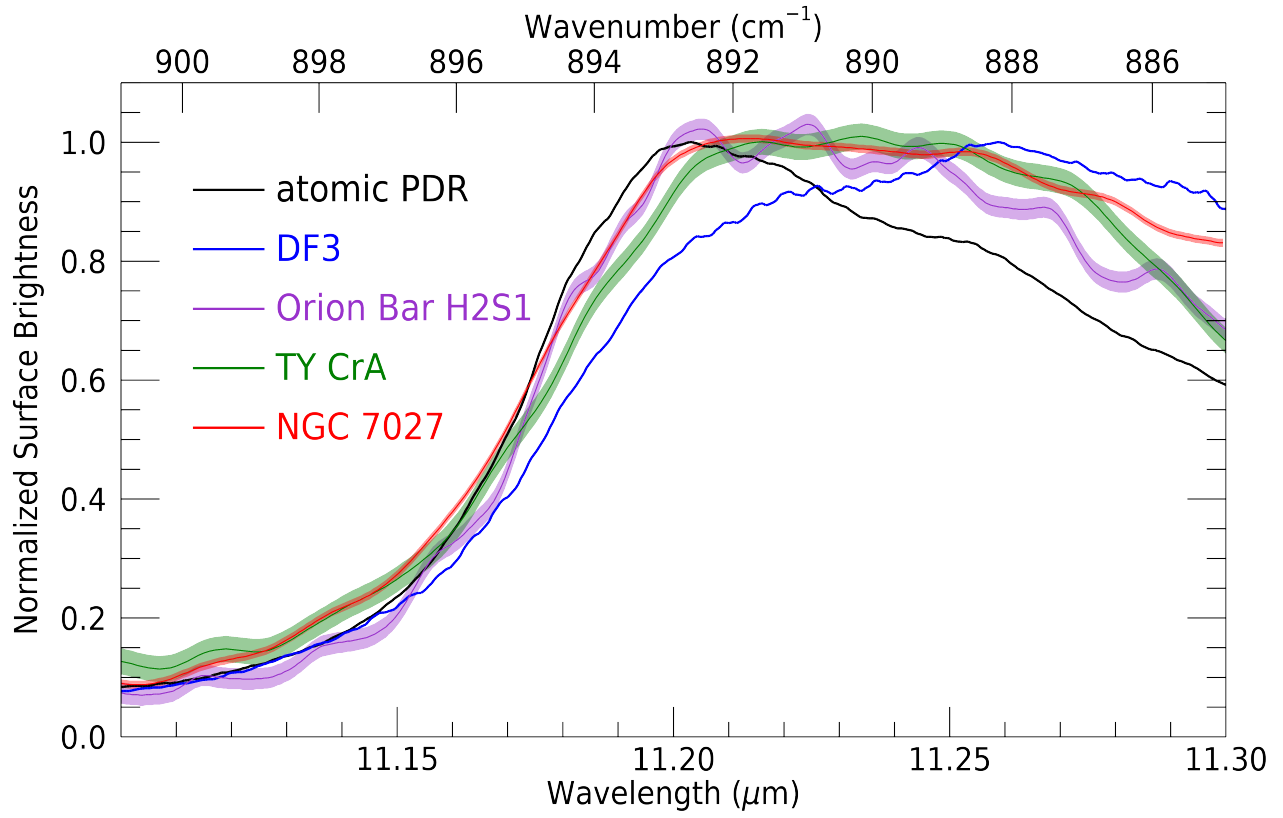}  
    \caption{ Expanded view of the blue wing of the 11.2 \mic band.  
    The one sigma error bars of the \textit{ISO SWS} spectra (Orion Bar H2S1, TY CrA, and NGC 7027) are shown as shaded regions on either side of the profiles. 
    The error bars for the \textit{JWST} profiles of the atomic PDR and DF3 regions are too small to be seen, but uncorrected residual fringes are present.  }
    \label{fig:expanded 11.2}
\end{figure}

Figure \ref{fig:expanded 11.2} shows an expanded view of 11.2 \mic blue wing. 
Excluding the DF3 spectrum, which is thought to have two components (see Section 2.4), the average wavelength at half peak intensity of the blue wing, 
$\lambda_B (11.2)_{1/2}$ is 11.171 \mic $\pm$ 0.002 \mic (895.21 $\pm$ 0.16 cm$^{-1}$). 
The uncertainty is the spread in the wavelength or frequency at the half-power point. The value of \lblue is identical to the result obtained by \citet{Candian15}. 
For the atomic PDR spectrum alone, we find $\lambda_B (11.2)_{1/2}$ = 11.169 \micron. 
Given the very small dispersion of $\pm$ 0.002 \micron, only a small number of gas phase PAH species can match the value of $\lambda_B (11.2)_{1/2}$.  
This is discussed further in Section \ref{sec:constant 11.2 AIB}.

\subsection{The red wing of the 11.2 \texorpdfstring{$\mu$m}{micron} AIB}  
\label{sec:11.2 red wing}

There are two explanations for the variable red wing of the 11.2 \mic AIB.  
\citet{Mackie21, Mackie22} show that, for a single PAH molecule, the steepness of the blue wing is sensitive only to the rotational temperature of the PAH while the red wing is affected by both the energy of the absorbed UV photon and the size of the PAH. 
Alternatively, Khan et al. (2025) propose that the 11.2 \mic AIB consists of two components, one at 11.207 \mic and the other at 11.25 \micron. 
The intensity of 11.25 \mic component relative to the 11.207 \mic is variable, and this gives rise to the variations in the red wing of the 11.2 \mic AIB.  
The 11.25 \mic band is attributed to VSGs (PAH clusters), and it is strongest in the UV-shielded DF3 region. 
The spectrum of the atomic PDR region lacks the 11.25 \mic component and thus has the narrowest FWHM.

\subsection{The 6.2, 7.7, and 8.6 \texorpdfstring{$\mu$m}{micron} AIB profiles}  
\label{sec:6.2 7.7 8.6}

The profiles of the 6.2, 7.7, and 8.6 \mic bands have been extensively discussed by \citet{Peeters02}, \citet{vanDiedenhoven04}, and \citet{Chown24},
and we briefly discuss in this section the band profiles in connection to the objects shown in Figures 1-4.  These bands originate from different vibrational modes: the 6.2 \mic band arises from the CC stretching mode, the 7.7 \mic band from mixed CC stretching and CH in-plane bending modes, and the 8.6 \mic band from the CH in-plane bending with some CC stretching modes \citep{Allamandola89, Chown24}.
Figures 5, 6, and 7 present the continuum-subtracted AIB spectra of selected sources from the \iso archive and \textit{JWST}.  
The continuum was removed as described by \citet{Peeters02}.
Within the PAH hypothesis, the 6.2, 7.7, and 8.6 \mic bands are attributed to vibrational modes of PAH cations \citep{Allamandola89, Allamandola99, Tielens08}. 

\begin{figure}
\includegraphics[width=\columnwidth]{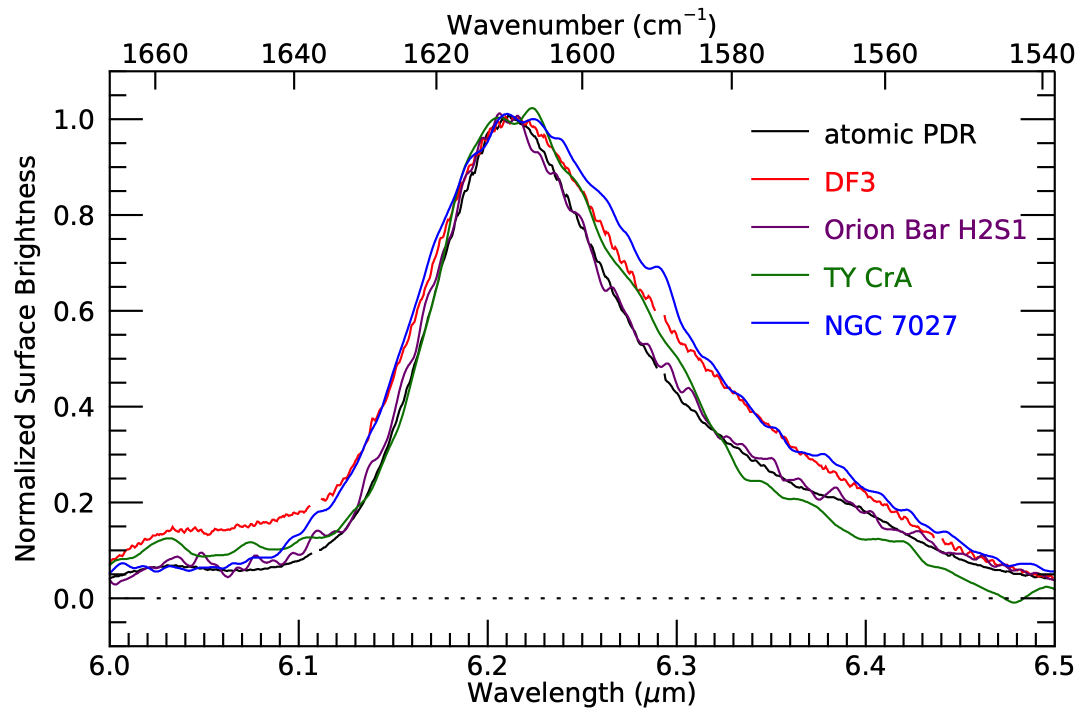} 
\caption{Comparison of the 6.2 \mic AIB. The continuum was removed as described by \citet{Peeters02}. 
The peak wavelength is nearly the same in all of the sources. Although the FWHM varies, it is no wider than that of DF3.  
}
\label{fig:6.2 comparison}
\end{figure}

Figure \ref{fig:6.2 comparison} shows the comparison of the 6.2 \mic AIB profiles.  
The peak wavelength is nearly constant within the uncertainties, but the FWHM of the DF3 and NGC 7027 profiles are broader than the atomic PDR profile. 
The systematic noise in the ISO spectrum of NGC 7027 is relatively large and so it is not possible to say definitely if the red wing is the same for DF3 and NGC 7027. 
The 7.7 and 8.6 \mic AIBs form a complex of overlapping bands \citep{Peeters02, vanDiedenhoven04}.
The 7.7 \mic band consists of two main subbands at 7.6 and 7.8 \micron, and these vary in relative strength in the transition from Class A to Class B \citep{Peeters02, Pino08}.

\begin{figure}
\includegraphics[width=\columnwidth]{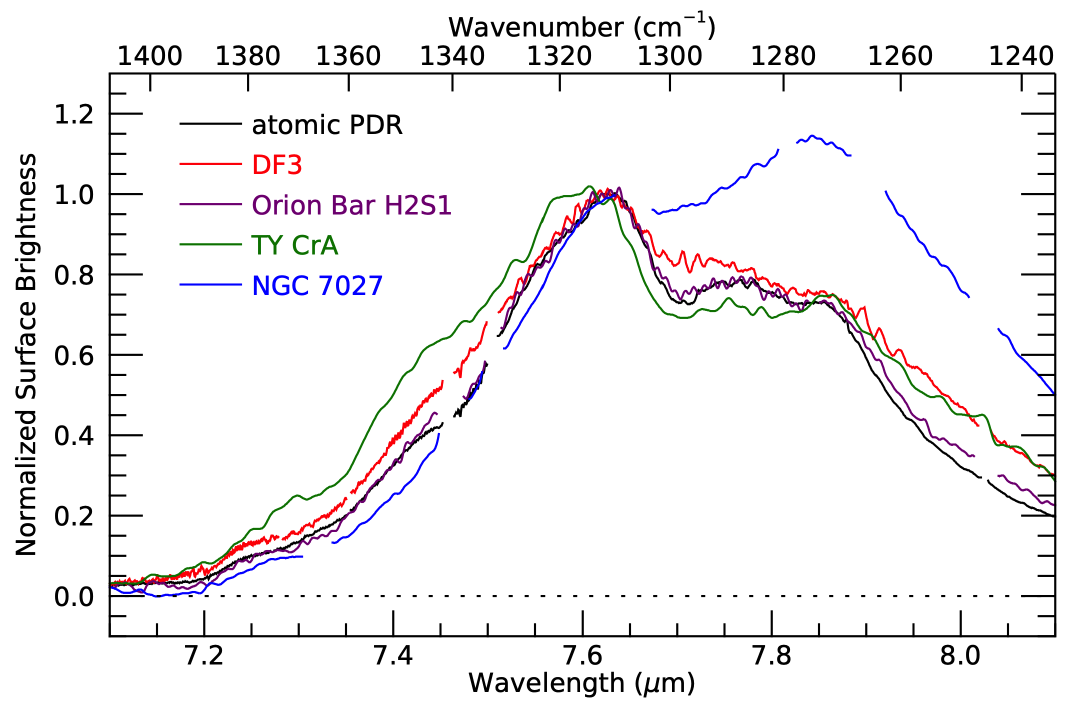} 
\caption{ Comparison of the 7.7 \mic AIB. The continuum was removed as described by \citet{Peeters02}. 
This AIB has two main peaks at 7.6 and 7.8 \mic.  The \textit{ISO SWS} spectrum of TY CrA appears to be shifted and this needs to be confirmed because of the higher noise and systematic errors.  
The spectrum of NGC 7027 is a Class B type and we show it to illustrate that some sources can be Class A at 3.3 \mic but Class B in the mid-IR.
}
\label{fig:7.7 comparison} 
\end{figure}

\begin{figure}
\includegraphics[width=\columnwidth]{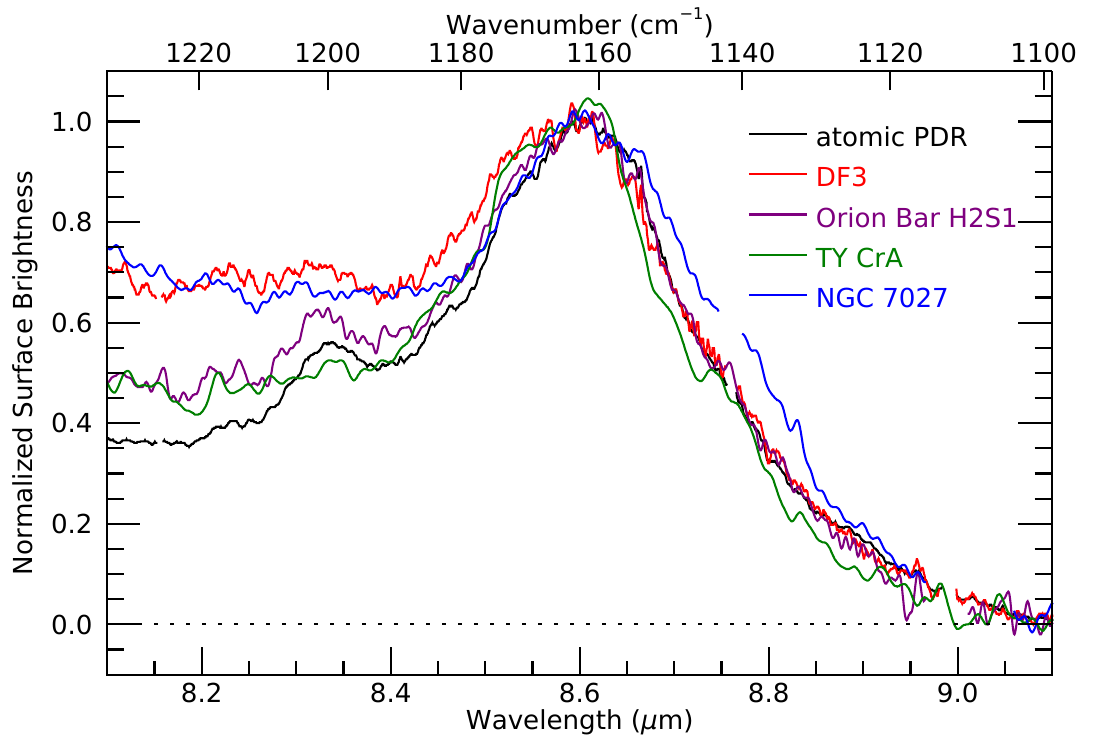} 
\caption{ Comparison of the 8.6 \mic AIB.  The continuum was removed as described by \citet{Peeters02}.
The peak wavelength is nearly the same in all sources.  
The FWHM is wider for the DF3 and NGC 7027 spectra compared to the atomic PDR.  
}
\label{fig:8.6 comparison} 
\end{figure}

Figure \ref{fig:7.7 comparison} shows the comparison of the 7.7 \mic AIBs. 
There is more variation in the band profiles than observed in the 3.3, 6.2 and 11.2 \mic AIBs.
Figure \ref{fig:8.6 comparison} shows the comparison of the normalized 8.6 \mic AIB. 
When normalized either at 7.6 \mic AIB peak or at the 8.6 \mic AIB peak the atomic PDR has the narrowest FWHM and the DF3 the broadest for the Class A sources. 
Interestingly, NGC 7027 has a Class B spectrum even though it has a Class A 3.3 \mic AIB with the same \lred as other Class A sources.  
The mixed Class A and B for the AIBs is observed in planetary nebulae \citep{vanDiedenhoven04},
and this may be the result of combining warm and cool dust emission components within the large \iso aperture.

In summary, the constant peak wavelengths of the 6.2, 7.7, and 8.6 \mic AIBs add more constraints that any proposed set of PAHs and their cations must satisfy. 
A comprehensive description of the Class A 6.2, 7.7, and 8.6 \mic bands is given by \citet{Chown24}.

\section{Unresolved questions with the PAH hypothesis}   
\label{sec:unresolved questions}

In this section we discuss the lack of diversity of the AIBs, in particular the nearly constant \lred and \lblue, and how this implies a limited number of PAHs giving rise to the 3.3 \mic AIB and a separate limited number of PAHs giving rise to the 11.2 \mic AIB.
We also discuss the implications for 11.2/3.3 band ratios and spectral features expected to be present in the UV and optical wavelength ranges.

\subsection{The constant AIB peak wavelength of Class A sources}   
\label{sec:constant AIB peak}

The AIB peak wavelength are constant in Class A sources \citep{Peeters02,vanDiedenhoven04}.  
The band profiles are also similar but there are variations in the FWHM.  This can be seen in Figures \ref{fig:3.3 profile}, \ref{fig:11.2 profile}, \ref{fig:6.2 comparison}, \ref{fig:7.7 comparison}, and \ref{fig:8.6 comparison}.
Fitting of the AIB bands using the Ames PAHdb has led to the hypothesis that there may be family of large symmetric compact PAHs referred to as the "grandPAHs" \citep{Andrews15, Peeters17, Bouwman19, Chown24}.
These are the PAHs that have survived a “weeding” out process that destroys less robust PAH species by the strong UV irradiation in the ISM. 
However, key details are undefined such as the definition for a PAH family, the molecular species in the PAH family, and the minimum number of carbon atoms ($N_C$) that defines a large PAH.   
Overall, the consistent AIB peak wavelengths suggest that a stable and limited number of PAH molecules is present in many sources and in many excitation conditions.  
As discussed in Sections 3.2, 3.3, and 3.4, the nearly constant \lred and \lblue require \textit{very specific} small and large PAHs.

\subsection{The 3.3 \texorpdfstring{\mic}{mic}AIB profile and the constant \textbf{ \texorpdfstring{$\lambda_R (3.3)_{1/2}$}{lambda\_R (3.3)\_1/2} } }   
\label{sec:3.3 AIB profile}

Small PAHs in the range \nc $\sim 20-40$ are needed to fit the AIB spectrum  \citep[i.e.][]{Verstraete01, Pech02, Andrews15, Bauschlicher18, Ricca24}.
In addition, models fitting the AIB spectrum with PAH nanoparticles adopt 30 carbon atoms as the lower limit for the size of the nanoparticles \citep{Draine01, Hensley23, Hirashita23}.
Since experiments and theory suggest that PAHs with fewer than 30 carbon atoms do not survive in the ISM \citep{Jochims94, LePage03}, we define “small PAHs” as those having $30-50$ carbon atoms.  
These small PAHs are the primary contributors to the emission in the 3.3 \mic AIB.  

\citet{Joblin95-1} showed the temperature dependence of the PAH peak wavelength and FWHM with laboratory spectra of gas phase PAHs with \nc $\leq$ 32. 
In the case of ovalene, C$_{32}$H$_{14}$, an increase in temperature of 800 to 900 K results in a redshift of 4 cm$^{-1}$ in the peak wavelength and a broadening of the FWHM of 5 cm$^{-1}$. 
This is significantly larger than the spread in \lred of 1.1 cm$^{-1}$ which was seen for the variety of sources shown in Figure 1. 
Thus, temperature changes as little as 50 K or less are discernible as shifts in the peak wavelength and FWHM in small PAHs. 
The nearly constant 3.3 \mic AIB profile and \lred imply that the PAH species and their excitation temperature are remarkably uniform in different excitation environments.

With the exception of \citet{Verstraete01} and \citet{Pech02}, models of the AIB emission using the Ames PAHdb do not take into account the temperature-dependent shift in the peak wavelength and change in the FWHM. 
The Ames PAHdb uses a constant redshift of the peak wavelength and a constant FWHM for all PAHs.  
This is understandable as the laboratory data does not exist to take into account the temperature dependence for the large PAHs (\nc > 50).  
Accurate modeling of the 3.3 \mic AIB profile requires anharmonic density functional theory (DFT) calculations\footnote{The NASA Ames PAH database uses a harmonic potential and does not account for the overtone or Fermi resonance bands. Therefore only qualitative results can be obtained for the 3.3 \mic AIB.  However, the Ames PAH database is a necessary spectroscopic tool for PAHs with more than 32 carbon atoms due to the lack of laboratory spectra of large PAHs. Anharmonic DFT calculations are necessary for the 3 \mic spectral region but it is not capable of computing spectra of PAHs with more than 22 carbon atoms \citep{Lemmens21}.
}, but this is limited to PAHs with $\leq 18-22$ carbon atoms \citep{Mackie18, Lemmens21}.
For the small PAHs, the laboratory spectra of \citet{Joblin95-1} could be used.  
For the large PAHs, \citet{Tokunaga21} use the \citet{Joblin95-1} to extrapolate the temperature dependence of the peak wavelength and FWHM but the uncertainties are large.

There is additional compelling evidence suggesting that the same PAH species are responsible for the AIB emission regardless of the source or excitation environment. 
\citet{Ricca24} noted that the two components of the 3.3 \mic AIB, 3.246 and 3.29 \micron, have a 3.246/3.29 band ratio that is the same across different regions of the Orion Bar. 
They conclude from this that these emission bands arise from the same set of PAHs.  
\citet{Pasquini24} find their cluster analysis indicates that the $3.4-3.5$ \mic emission bands arise from a single sidegroup that is attached to similar-sized PAHs. 
This reinforces the idea that the PAH species that give rise to the Class A 3.3 \mic AIB do not change under many different excitation environments.  
While these studies are limited to the Orion Bar, the consistency of the 3.3 \mic AIB profile in many sources (examples of which are shown in Figure 1) strongly implies that the same set of PAHs exist everywhere in the ISM. 

Laboratory data and anharmonic DFT calculations show a 3.3 \mic band profile that has a steep blue wing and a broader red wing, a profile that is the opposite of the observed 3.3 mic AIB profile which has a broad blue wing and a steeper red wing.  
We can see this clearly for naphthalene in Figure 2 of \citet{Mackie22}. 
Further evidence comes from \citet{Mackie18}, who present anharmonic DFT cascade emission spectra for small PAHs at excitation energies of 3, 5, 7, and 9 eV. 
In their supplemental material, the steep blue wing is consistently seen at excitation energies $\geq$ 5 eV for anthracene (C$_{14}$H$_{10}$), benz[a]anthracene (C$_{18}$H$_{12}$), chrysene (C$_{18}$H$_{12}$), naphthalene (C$_{10}$H$_{8}$), phenanthrene (C$_{14}$H$_{10}$), pyrene (C$_{16}$H$_{10}$), and tetracene (C$_{18}$H$_{12}$).  
Similarly, \citet{Esposito24-2} show that the anharmonic cascade emission spectrum of indene (C$_9$H$_8$) also has a steep blue wing.  
The discrepancy between these calculated profiles and the observed 
3.3 \mic AIB profile is presently unexplained.  

In summary, the emission in the 3.3 \mic AIB is dominated by a limited number of small PAHs which can reach the high temperature necessary to explain the 3.3 \mic AIB. Currently, there is no information about which small PAH species could explain the nearly constant $\lambda_R (3.3)_{1/2}$, as anharmonic DFT calculations of PAHs with $30-50$ carbon atoms are not yet possible. Nevertheless, the uniformity of the 3.3 \mic AIB profile and the nearly constant \lred suggest that the same set of small PAH species is responsible for the 3.3 \mic AIB. This consistency is particularly notable given that \lred is very sensitive to the PAH species and the excitation temperature.
In addition we note that the observed profile of the 3.3 \mic AIB is not consistent with the calculated profile.  

\subsection{The constant \textbf{ \texorpdfstring{$\lambda_B (11.2)_{1/2}$}{lambda\_B (11.2)\_1/2} } and large PAHs }  
\label{sec:constant 11.2 AIB}

\citet{Candian15} modeled the nearly constant $\lambda_B (11.2)_{1/2}$. 
They found that the nearly constant blue wing could be explained by two large PAHs such as C$_{54}$H$_{18}$ and C$_{66}$H$_{20}$ and if the 11.2 \mic band reaches an asymptotic limit in wavelength as the PAHs get larger. 
Figure 8 presents the 11.2 \mic solo C-H out-of-plane (OOP) absorption wavelengths of symmetric and compact PAHs that were obtained from the Ames PAHdb. 
The PAH molecules shown are compact and symmetric PAHs that were selected from those discussed by \citet{Weisman03, Bauschlicher08, Ricca12}.
In order to focus only on the intrinsic spread of the absorption wavelengths, no frequency shifts were applied to the frequencies obtained from the Ames PAHdb. 

\begin{figure}
	\includegraphics[width=\columnwidth]{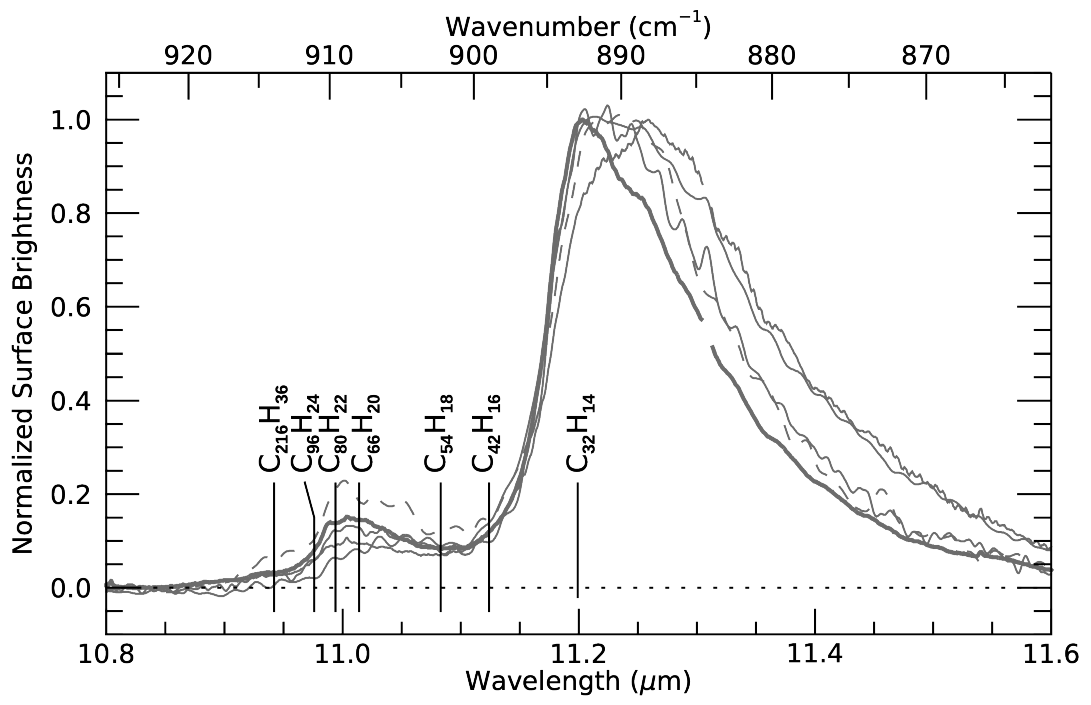}  
    \caption{ Spectra from Figure 3 compared to the absorption wavelengths  of symmetric and compact PAHs calculated from the NASA Ames PAH Database (indicated by the vertical lines).  
    No corrections were made for anharmonic and temperature shifts.
    An asymptotic limit of decreasing wavelength with increasing PAH size is reached with $C_{216}H_{36}$.  }
    \label{fig:PAH 11.2 wavelengths}
\end{figure}

As shown in Figure 8, \lblue strongly dependent on the PAH species. 
The precise \lblue in many different objects and physical conditions suggests a very specific PAH or set of PAH molecules that give rise to the 11.2 \mic AIB.  
\citet{Mackie21, Mackie22} show that for a single PAH, \lblue closely matches the absorption peak at 0 K and that it is largely unaffected by the excitation conditions. 
Figure \ref{fig:PAH 11.2 wavelengths} shows that the absorption wavelength varies by $\sim$0.14 \mic ($\sim$12 cm$^{-1}$) between C$_{54}$H$_{18}$ to C$_{216}$H$_{36}$. 
This is in contrast to the extremely small variation in \lblue of about $\pm$ 0.002 \mic (0.16 cm$^{-1}$) reported in Section \ref{sec:11.2 blue wing}.  
In addition, \citet{Sadjadi15} was not able to fit the 11.2 \mic AIB band with pure PAH molecules.

\begin{figure}
	\includegraphics[width=\columnwidth]{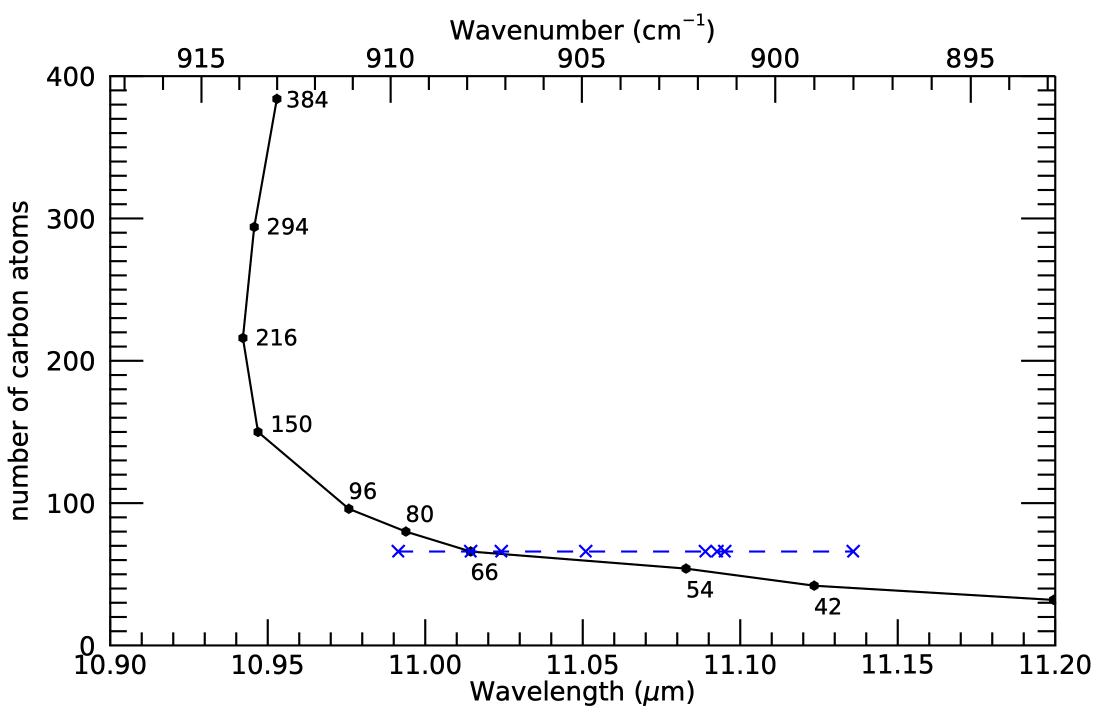}  
    \caption{The 11.2 \mic OOP absorption wavelength of symmetric and compact PAHs that were calculated using the Ames PAHdb. The wavelength decreases with increasing PAH size and reaches a limit at 216 carbon atoms.  The PAHs were selected from \citet{Weisman03}, \citet{Bauschlicher08}, and \citet{Ricca12} and consists of C$_{32}$H$_{14}$, C$_{42}$H$_{16}$, C$_{54}$H$_{18}$, C$_{66}$H$_{20}$, C$_{80}$H$_{22}$, C$_{96}$H$_{24}$, C$_{150}$H$_{30}$, C$_{216}$H$_{36}$, C$_{294}$H$_{42}$, and C$_{384}$H$_{48}$.
    ‘X’ symbols show the wavelength of the C$_{66}$H$_{20}$ isomers.
    }
    \label{fig:11.2 asymptotic limit}
\end{figure}

Figure \ref{fig:11.2 asymptotic limit} shows a plot of the OOP absorption band wavelength as a function of PAH size for compact and symmetric PAHs. 
As the PAH size increases, the OOP band wavelength decreases and approaches an asymptotic limit at about 150 carbon atoms.  
Although Figure \ref{fig:11.2 asymptotic limit} suggests a "solid state" limit alluded to by Pech et al. (2002), PAH isomers will produce a large scatter in the wavelengths as shown for C$_{66}$H$_{20}$. 
Figure \ref{fig:11.2 asymptotic limit} includes only the C$_{66}$H$_{20}$  isomers given in the Ames PAHdb, and the scatter would very likely be larger if all of the many isomers of C$_{66}$H$_{20}$ were included.  
The isomers creates great difficulty in seeing how a small number of PAHs could explain the blue wing of the 11.2 \mic AIB since only those isomers that match \lblue would be expected to survive the weeding out process in the ISM.  
This seems very unlikely. 
In the case of C$_{66}$H$_{20}$, for example, the zero point energy of all of the isomers given in the Ames PAHdb differ at the most by only 5.0 kcal/mole, or 0.22 eV. Thus survivability of all isomers against photodissociation is about the same.

Constraints on the type of PAHs that could match the OOP bands at $11-13$ \mic have been explored by \citet{Hony01} and \citet{Ricca24}.  
They find that PAHs described as zigzag and armchair by \citet{Ricca24} fit best.  
The PAHs illustrated by \citet{Hony01} in their Figure 9 are not consistent with compact symmetric PAHs.

In summary, the observations show an extremely constant $\lambda_B (11.2)_{1/2}$. 
If the constant value of \lblue is attributed to a limited number of compact, symmetric PAHs, then their isomers must also be considered.  
The selective formation and survival of only very specific PAHs and the right set of isomers in the ISM presents a significant challenge for the PAH hypothesis.

\subsection{The 11.2/3.3 band ratio and PAH sizes}    
\label{sec:11.2/3.3}

The 11.2/3.3 band ratio commonly is used to estimate PAH sizes.  Observational data show that the 11.2/3.3 band ratio is typically in the range of 1.2 to 4. 
Modeling based on this ratio yields PAH sizes in the range of $40-183$ carbon atoms, depending on the object and proximity to the exciting source \citep{Ricca12, Croiset16, Maragkoudakis20, Knight21, Knight22}.  
However, the reliability of the 11.2/3.3 ratio as an indicator of the PAH size warrants re-examination, given the limited number of PAH species implied by \lred and $\lambda_B (11.2)_{1/2}$.   
The estimates of the PAH sizes in these studies use the Ames PAHdb to calculate the 11.2/3.3 ratio as a function of the number of carbon atoms in the PAH molecule. 
The emission from the 3.3 \mic AIB decreases exponentially and becomes negligible when \nc $\gtrsim$ $80-100$ whereas the emission from the 11.2 \mic AIB is mostly emitted by PAHs with \nc $\gtrsim$ 80 (Schutte et al. 1993, see their Figure 9).  

Therefore, the observed ratio in the AIBs reflects contributions from two distinct PAH populations:  the smaller PAHs ($\sim 30-50$ carbon atoms) that contribute most of the 3.3 \mic AIB emission and the larger PAHs ($\gtrsim$ 80) that contribute most of the 11.2 \mic AIB emission. There must be two populations, since if there were only one population of PAHs, then individual PAH molecules would have to simultaneously match \lred and $\lambda_B (11.2)_{1/2}$.  This would imply a specific PAH molecule as a universal carrier of the AIB, which is improbable.  Variations in the 11.2/3.3 band ratio may simply reflect the relative abundance of the small and large PAHs. However, the plausibility of having only a limited number of PAH molecules is questionable as discussed in Sections 3.5 and 3.6.

\subsection{The UV and optical absorption bands of PAHs}   
\label{sec:UV-optical absorption}

PAHs typically show strong absorption at 300-400 nm and have been extensively studied in the laboratory \citep[i.e.][]{Salama96, Ruiterkamp02, Salama11, Steglich11, Zhen16}. 
In addition, DFT calculations by \citet{Weisman03} show that strong and narrow absorption bands are expected at 400-700 nm from compact symmetric PAHs.
Thus far, searches for these strong absorption bands in this wavelength range have been unsuccessful \citep{Clayton03, Salama11, Gredel11, Steglich11, Steglich12}.  
This has led to the conclusion that either the gas phase PAHs do not exist in sufficient abundance or there are so many PAHs that individual PAH absorption features are too weak to be detectable.  
The lack of spectral diversity and the narrow range of $\lambda_R (3.3)_{1/2}$ and $\lambda_B (11.2)_{1/2}$ implies a small number of PAHs and this causes difficulties with the latter idea.

To illustrate the problem, we start with the estimate of the fraction of interstellar C atoms that are in PAHs, $\sim 3.5 \times 10^{-2}$ (Tielens 2008).  He estimates a PAH fractional abundance of N(PAH)/N(H) $\sim 3 \times 10^{-7}$ for the case where all of the PAH molecules have 50 carbon atoms.  If we have only “a few” PAHs giving rise to the AIBs and if we assume “a few” means $3-5$ molecules, then we get the estimates for the fractional abundance of PAHs where each species has equal abundance as shown in Table 1.

\begin{table}
    \centering
    \begin{tabular}{|c|c|c|}
        \hline
            & 1 PAH species  & 3-5 PAH species  \\
        \nc per PAH species  &  N(PAH)/N(H) x10\textsuperscript{-7} & N(PAH)/N(H) x10\textsuperscript{-7} \\
        \hline
        30 & 5 & 1.0-1.7 \\
        50 & 3 & 0.6-1.0 \\
        75 & 2.2 & 0.4-0.7 \\
        100 & 1.5 & 0.3-0.5 \\
        \hline
    \end{tabular}
    \caption{Fractional abundances assuming 1 PAH species and 3-5 PAH species}
    \label{tab:pah_species}
\end{table}

Thus we see that for $3-5$ PAH species, the PAH fractional abundance is $\gtrsim 3 \times 10^{-8}$ for up to 5 PAH species with 100 carbon atoms or less.  
However, \citet{Gredel11} measured the absolute absorption cross-sections for six PAHs with $14-42$ carbon atoms at $3050-3850$ \AA.  
With observations from the Very Large Telescope (VLT), they derived upper limits for the PAH fractional abundance of $< 0.2 - 2 \times 10^{-9}$ for all but one of the PAH molecules they studied.
Therefore, if only a few PAH species were present in the ISM, then the observed upper limits are about 10 to 100 times below what is expected.

The lack of spectral diversity and the narrow range of \lred and \lblue implies a limited number of PAHs, 
and the absence of detectable PAH absorption features at UV and optical wavelengths further challenges the validity of the PAH hypothesis.

\subsection{The 217 nm absorption feature}
\label{sec:217 nm absorption}

The carrier of the 217 nm absorption feature remains unidentified but it may be a graphitic material, gas phase PAHs, carbonaceous material such as hydrogenated amorphous carbon (HAC), or PAH clusters \citep[i.e.][]{Joblin95-1, Mennella96, Schnaiter98, Duley12, Gavilan16, Herrero22, Bernstein24}.
Since 10-20\% of the infrared luminosity of galaxies are emitted in the AIBs, a connection between the 217 nm feature and the AIBs is likely. Within the framework of the PAH hypothesis, a large number of gas phase PAHs is needed to obtain an approximate fit to the 217 nm absorption feature and simultaneously achieve a relatively smooth absorption profile 
\citep{Joblin92, Malloci04, Steglich11, Steglich12, Lin23}.
This is inconsistent with a limited number of gas phase PAHs.

\section{Other bands and the continuum emission}  
\label{sec:other bands and the continuum}

In the astronomy literature, the term "PAH"  includes a broader range of molecules than the strict definition in chemistry. 
It includes molecules with side groups such methyl (CH$_3$) or additional H atoms, heterosubstituted PAHs (molecules containing nitrogen or oxygen),  dehydrogenation (carbon ring molecules with some or all H atoms removed), and PAH clusters (aggregates of PAH molecules and often considered to be VSGs) \citep{Peeters21}.  
In this section, we briefly discuss some of these aspects of the PAH hypothesis and the unresolved issues.  

\subsection{The aliphatic emission bands at 3.4 \texorpdfstring{$\mu$m}{micron} }
\label{aliphatic bands}

The 3.4 \mic emission complex consist of a moderately strong bands at 3.40, 3.465, 3.516 \mic, with the 3.40 \mic band being a blend of two closely spaced bands \citep{Chown24}. 
These emission bands are attributed to aliphatic side-groups such as -CH$_3$ attached to PAHs \citep{Joblin95-2, Pauzat99, Yang16, Bouteraon19, Buragohain20} or superhydrogenated PAHs 
\citep{Wagner00, Materese17, Jensen22}. 
It also could be a mixture of both types of aliphatic bonds \citep{Wagner00, Mackie18-2}.  
\citet{Esposito24-2} find that there can be a significant aromatic contribution to the bands at 3.4 \micron, although this conclusion is based on indene and 2-ethynyltoluene, molecules with only 9 carbon atoms.
Significantly, the 3.4 \mic emission band peaks are consistent across Class A sources, underscoring the lack of spectral diversity noted for the AIBs. 

The 3.4 \mic emission bands are almost always associated with the 3.3 \mic AIB.  It is never observed without the 3.3 \mic AIB.  This strong correlation suggests that both features originate from the same molecular carrier. As mentioned in Sec. 3.2, \citet{Pasquini24} find evidence that the 3.4 \mic emission bands arise from a single sidegroup that is attached to similar-sized PAHs from their cluster analysis.  Although the 3.4 \mic aliphatic bands contain 2 to $\sim$12\% of the carbon atoms in the carrier \citep{Yang23} depending on the source, their consistent presence and profile across Class A sources suggests a molecular homogeneity of the carrier and the side groups in diverse astrophysical environments.
        
\subsection{The plateau emission}
\label{sec:plateau emission}

Broad emission features underlying the AIBs at 3.4, 8, 12, and 17 \mic are referred to as the “plateau” emission features \citep[e.g.][]{Kwok13, Peeters21}.  
In the context of the PAH hypothesis, the origin of these features are assumed to be PAH clusters \citep[i.e.][]{Tielens08} or VSGs \citep[i.e.][]{Joblin08}.
The plateau emission features are removed, along with the continuum, using spline fitting in order to isolate what are considered to be the true PAH emission bands \citep[e.g.][]{Hony01, Peeters02, Pilleri15}.  
This procedure is ad hoc since there is no experimental or theoretical basis for assuming the plateau emission is a distinct component separate from the AIBs. 
An exception is the 3.4 \mic plateau emission which \citet{Wagner00} assigns to combination bands.  
For the 6-15 \mic AIBs, \citet{Boulanger98} uses a Lorentzian profile to fit the AIBs without assuming there is a separate plateau emission component.  
Similarly, \citet{Smith07} uses a Drude profile without assuming a plateau component.
The validity of assigning the plateau emission to PAH clusters or VSGs requires more rigorous examination.

\subsection{Other emission bands}
\label{sec:other bands}

Moderately weak to moderately strong emission bands have been observed at 5.25, 5.75, 12.0, 12.7, 13.5, 15.8, 16.4, 17.4, and 17.8 \mic \citep{Peeters12, Kwok22, Chown24}, along with numerous weaker emission features \citep{Chown24}.  
All of these bands are thought to be PAH emission bands, but there is no definitive identification to specific PAH molecules.  
Note that the 17.4 \mic band is a blend of a PAH band and C$_{60}$ \citep{Sellgren10}.
There is clearly much work ahead to identify these emission bands.
        
\subsection{The continuum emission}

Within the PAH hypothesis, the 1-5 \mic continuum emission is referred to as a quasi-continuum that arises from highly vibrationally excited molecules and is composed of many weak bands \citep{Allamandola89, Allamandola21, Esposito24}.  
In this scenario a flat emission spectrum is produced.  
However, anharmonic DFT calculations by \citet{Esposito24} suggest that achieving a smooth quasi-continuum requires contributions from a large number of PAH species. 
This idea seems unlikely to be viable if the number of PAH molecules is small.  
Alternatively, \citet{Lacinbala23} proposes that recurrent fluorescence from carbon clusters could be an explanation for the 1-5 \mic continuum emission.

\section{Discussion}
\label{sec:discussion}

To date, the PAH hypothesis remains the most widely accepted explanation for the general properties of the AIBs – its spectral characteristics, the physics of excitation and emission, and the observed variations of the band ratios\footnote{Other ideas for the AIB carrier are listed in Appendix A.}.
The AIB modeling demonstrates that the AIB emission arises from a carrier material with chemical bonds characteristic of PAH molecules. 
Further support comes from radio detections of PAHs and aromatic nitriles in dark clouds \citep[i.e.][]{Burkhardt21,McGuire21,Wenzel25b,Wenzel25a}.
These detections provide direct observational evidence for PAH-related species in space. 
Nevertheless, there are fundamental issues pointed out in this paper that need to be addressed before the PAH hypothesis can be considered definitively established.

The set of PAH molecules giving rise to the 3.3 \mic and 11.2\mic AIB is very consistent whether it is "fresh" material produced in planetary nebula or is highly processed material in HII regions, and this point warrants more attention. 
The objects in Figures \ref{fig:3.3 profile} and \ref{fig:11.2 profile} span a wide range of excitation conditions, with the interstellar radiation field ranging from 2.6$\times$10$^3$ to 6$\times$10$^5$ in Habing units\footnote{1.6$\times$10$^{-3}$ erg cm$^{-2}$ s$^{-1}$;  \citet{Habing68}.} with a wide range of spectral energy distribution in the UV \citep[][]{Peeters02, Tokunaga21}.
Despite these differences, the nearly constant \lred and \lblue suggests that only a limited number of gas phase PAH species are responsible, one set to explain the 3.3 \mic AIB and another set to explain the 11.2 \mic AIB.  

The Class A spectrum, which is the focus of this paper, is characteristic of Galactic sources as well as normal and starburst galaxies \citep[i.e.][]{Smith07}. 
\cite{Xie18} found that the same Class A spectral template can be used to model the AIB emission in high latitude clouds, normal galaxies, active galaxies, quasars, and ultraluminous infrared galaxies. Thus in the PAH hypothesis, the AIB carrier is a stable, reproducible collection of a surprisingly modest number of PAH molecules that are universally produced throughout the Galaxy and beyond.

Why is it that only a small number of distinct, well-defined PAH species appear to form and persist in the ISM? We know that PAHs are formed in planetary nebulae, but they are processed in the interstellar medium by shocks and chemistry. 
\citet{Shannon19} and \citet{Goicoechea25} describe two formation pathways:  bottom up, where PAHs form from simple molecules and radicals, and top-down, where PAHs result from fragmentation or evaporation from VSGs (PAH clusters). 
These processes should produce a wide range of different gas-phase PAHs molecules in both the ISM and circumstellar environments. 
Indeed fitting of the 3-15 \mic spectrum of AIBs typically use 25 to 50 PAHs \citep{Bauschlicher18, Andrews15, Ricca24} but can be as high as 300 PAHs \citep{Maragkoudakis20}.
Thus there is a conflict between the theoretical diversity of PAHs that can be produced in the ISM and the lack of spectral diversity of the AIBs.

We do not have a quantitative way of estimating the size and composition of the PAH molecules required to explain the AIB spectrum. 
As shown in Figure \ref{fig:PAH 11.2 wavelengths} the constant $\lambda_B (11.2)_{1/2}$ suggests that the large PAHs might consist of a very specific size range. 
If compact, symmetric PAH molecules are the most likely to survive in the ISM, this further restricts the possible pool of viable PAH species. Similarly, the consistent profile and peak wavelength of the 3.3 \mic AIB is a strong constraint on the number of small PAH species that could be the carrier, potentially even to a single PAH species.


\section{Summary}
\label{sec:summary}

Although the PAH hypothesis has been very successful in explaining many aspects of the AIB spectrum, there are a number of open questions regarding the PAH hypothesis.
We hope this paper will stimulate further research addressing these unresolved questions. 


The key challenges to the PAH hypothesis are:

\begin{enumerate}

    \item Lack of spectral diversity.  ISO SWS and JWST spectra show a lack of spectral diversity of the Class A AIB profiles with a nearly constant \lred and \lblue in many sources with different excitation conditions.  Explaining a nearly constant \lred and \lblue requires two narrowly defined sets of PAH species, one emitting primarily at 3.3 and the other at 11.2 \mic, which has observational consequences as noted below.

    \item Conflict with current models for the AIB emission.  A small number of PAH species is inconsistent with current modeling of the AIBs which assumes that a wide range of PAH types and sizes are the carriers of the AIB.

    \item Absence of UV/optical absorption features.  A small number of PAH species would be expected to produce strong absorption bands in the 300-700 nm, but this is not observed.  Furthermore, it would be difficult to explain the 217 nm absorption band with only a small number of gas phase PAH species.

    \item Other bands and the continuum emission. We note unresolved issues that warrant further investigation, including the 3.4 \mic emission bands, the plateau features, and the continuum emission. 

    \item Synthesis and survival of unique PAH species.  In order to preserve the PAH hypothesis, it is necessary to explain how only a small number of gas phase PAHs can be consistently produced and preserved in the ISM in a wide range of astrophysical environments and excitation conditions.

\end{enumerate}

\section*{Acknowledgements}

We thank the PDRs4All team for the excellent archive of the \textit{JWST} Orion Bar spectra that made this work possible.
We also thank Roger Knacke and Walter Duley for helpful discussions, and an anonymous referee for constructive comments.
TO acknowledges the support by the Japan Society for the Promotion of Science (JSPS) KAKENHI Grant Number JP24K07087. 

This research used the NASA Ames PAH Database \citep{Bauschlicher18,Boersma14,Mattioda20} and 
the Astrophysics Data System, funded by NASA under Cooperative Agreement 80NSSC21M00561.  
The Gemini AI 2.5 and Microsoft CoPilot tools were used to conduct literature searches and for editing sections of the paper for improving clarity.  
The content and meaning of the text was preserved in the use of the AI tools. All scientific content, interpretations, and conclusions are solely those of the authors.

\section*{Data Availability}


The data underlying this article will be shared upon request to the corresponding author.

\newpage


\bibliographystyle{mnras}
\bibliography{challenges_paper} 




\appendix

\section{Other ideas for the carrier of the AIB}

In addition to gas phase PAHs, various types of carbonaceous solids with PAHs have been proposed to be the carriers of the AIBs.  These include mixed aromatic/aliphatic organic nanoparticles \citep[MAONs;][]{Kwok11, Kwok13}, carbonaceous nano-grains \citep[][based on the \citet{Jones17} THEMIS model for dust grains]{Duley81, Elyajouri24}, laboratory analogs including amorphous carbon \citep{Borghesi87, Herlin98}, quenched carbonaceous composite \citep[QCC;][]{sakata87}, materials related to coal \citep{Papoular91}, chemical energy release in hydrogenated amorphous carbon \citep[HAC;][]{Duley11}, carbon nanoparticles \citep{Duley12}, materials related to petroleum extracts \citep{Cataldo13}, and disordered aromatic and aliphatic material \citep{Dartois20}.  

However, none of these ideas and laboratory analogs has been able to definitively link a specific material to the emission bands or to be able to explain the stochastic emission process that produces the AIBs.  
\citet{Kwok22} and \citet{Yang17a} summarize the issues related to identifying the carrier material.

Since the discovery of the first AIB \citep{Gillett73} over 50 years ago, the nature of the carrier of the AIBs continues to perplex us.  
Observations show that the Class A AIB is ubiquitous in the universe and that the carrier, containing aromatic chemical bonds, is gas phase PAHs, or carbonaceous dust particles, or both. 
The lack of spectral diversity points to a carrier that is very similar everywhere.
This should simplify the identification of a candidate material, 
but a conclusive identification remains elusive.
Nature continues to challenge our imagination and our understanding of astrochemistry.


\bsp	
\label{lastpage}
\end{document}